\documentstyle[preprint,aps]{revtex}
\begin{document}
\draft
\title {\bf Massive compact dwarf stars and C-field}
\author{L.P. Singh$^1$ and P.K. Sahu$^2$}
\address{$^{1}$Physics Department, Utkal University,
Bhubaneswar-751004, INDIA; E-mail: lps@utkal.ernet.in.} 
\address{$^{2}$Physical Research Laboratory, Theory Group,
Navarangpura, Ahmedabad 380 009, INDIA;
E-mail: pradip@prl.ernet.in.}
\maketitle
\begin{abstract}
The effect of C-field in high density matter has been studied.
We find that the negative energy and negative pressure of the
C-field helps in formation of massive compact stable neutron stars of
mass $\sim$ 0.5 solar mass which is in the range of 0.01 to 1.0
solar mass of recently observed dwarf stars.
\end{abstract}
There has recently been reported observations by gravitational
microlensing of dark dwarf stars in the mass range of
0.01-1.0$M_{\odot}$ (Alcock {\it et al.} 1993; Aubourg {\it et al.} 1993). 
Simple extrapolations of this observations have led people to speculate
(Boughn \& Uson 1995) that such bodies could be numerous and make up the 
``dark matter" inferred from the rates of rotation of galaxies
and superclusters of galaxies. In any case, this observation has
opened up the question of what these massive compact objects
themselves are. Cottingham, Kalafatis \& Vinh Mau (1994),
have proposed a model where these objects are identified as
quark stars formed after quark-hadron transition.
\par
We, in this work, suggest a different understanding of these
objects. The stability of these objects at unusual mass values
is indicative of a simultaneous increase in binding and reduction
in the internal pressure compared to the normal stars leading
to the balance between the two at lower mass values. Driven by
such an argument and since the creation field (C-field) of
Pryce, Hoyel \& Narlikar (1962) and Narlikar (1973) used in the context
of steady-state theory has exactly these characteristics of 
negative pressure and negative energy, we have tried to include
its contribution into the description of high density matter
interms of SU(2) chiral sigma model (Sahu, Basu \& Datta 1993).
\par
Eventhough, the hot big-bang model of the creation and evolution
of the Universe has gained acceptance over the steady-state 
cosmology, it does have problems associated with linearity of
Hubble flow and determination of the age of the Universe from 
the Hubble Space Telescope data (Narlikar 1993). It is, therefore, in
the fitness of things to explore the possibility of an explanation
for the problem at hand within the context of an alternative model
as has been attempted in other contexts (Weinberg 1972; Arp {\it et al.} 1990).
\par
The massless C-field was originally introduced by Pryce
and later extensively used by Hoyle \& Narlikar (1962) and Narlikar (1973) to
provide a field-theoretic understanding of the continuous creation
of matter in the steady-state evolution of the Universe preserving 
energy and momentum conservation. As a first approximation, we
take the C-fields to be noninteracting. Its energy-momentum
tensor is given by
\begin{equation}
T^{ik}=-f[C^iC^k-\frac{1}{2}g^{ik}C^lC_l]
\end{equation}
where $C_i$ is simply $\frac{\partial C}{\partial x^i}$ and f
$(>0)$ is a coupling constant. In fact, the energy density and
pressure contributed by C-field are given by
\begin{equation}
\epsilon = p = -\frac{1}{2}f\dot{C}^2
\end{equation}
where $\dot{C} \equiv \frac{\partial C}{\partial t}$. Again, if
all particles, with whose creation the C-field are
simultaneously created, have same mass m, then $\dot{C}=m$. On
quantisation of C-field, it has been argued that the C-field
does not lead to the usual problem of cascading associated with
negative energies once its effect of expanding the space-time
structure due to the feed back of energy-momentum tensor of C-field
on space-time geometry is taken into account. However, as is
borne out by our calculation, the negative energy of C-field
makes matter to condense to more compact size of mass of the
order reported above.
\par
We like here to state that the equation of state (EOS)of matter
above the nuclear matter 
density plays a crucial role to determine the equilibrium
structure of compact objects such as neutron stars. Upto about
nuclear density the equation of state is reasonably well known,
but the central density of compact objects can be almost an order
of magnitude higher. In this regime, the physics is unclear.
For that reason, attempt has been made to calculate EOS using
two major techniques: (i) non-relativistic and (ii) relativistic
field theories. The shape of baryonic potential is not known at
very small interparticle separations. Also, it is not clear that
the potential description will continue to remain valid at such
short ranges. Therefore, the non-relativistic approach may not
be adequate. In the recent times, the relativistic approach has
drawn considerable attention. In the relativistic approach, one
usually starts from a local, renormalizable field theory with
baryon and explicit meson degrees of freedom. The theory is
chosen to be renormalizable in order to fix the coupling
constants and the mass parameters by empirical properties of
nuclear matter at saturation. As a starting point, one chooses
the mean field approximation which should be reasonably good at
very high densities. This approach is currently used as a
reasonable way of parameterizing the EOS. However, in recent
years, the importance of the three-body forces in the EOS at
high densities has been emphasized by several authors (Jackson, Rho
\& Krotscheck 1985; Ainsworth {\it et al.} 1987).
This gives theoretical impetus to study the chiral sigma model,
because the non-linear terms in the chiral sigma Lagrangian can
give rise to the three-body forces.
\par
We also take the approach that the isoscalar vector field be
generated dynamically. Inclusion of such field is necessary to
ensure the saturation property of nuclear matter. The effective
nucleon mass then acquires a density dependence on both the
scalar and the vector fields, and must be obtained self-consistently. We
do this using mean-field theory wherein all the meson fields are
replaced by their uniform expectation values.
\par
The Lagrangian for an $SU(2)$ chiral sigma model that includes
an isoscalar vector field ($\omega_{\mu}$) is ($\hbar = 1 = c$)
(Sahu, Basu \& Datta 1993) and a non-interacting C-field is 
\begin{eqnarray}{\cal L} =
\frac{1}{2}\big(\partial_{\mu} \overrightarrow{\pi} .\partial ^{\mu}
\overrightarrow{\pi} + \partial_{\mu} \sigma\partial^{\mu} \sigma\big) -
\frac{\lambda}{4}\big(\overrightarrow{\pi} .\overrightarrow{\pi} +\sigma^{2} -
x^2_o\big)^2\nonumber\\ - \frac{1}{4} F_{\mu\nu} F_{\mu\nu} +
\frac{1}{2}{g_{\omega}}^{2}\big(\sigma^2 + \overrightarrow{\pi}^2\big)
\omega_{\mu}\omega^{\mu} \nonumber\\  + g_{\sigma} \bar{\psi} \big(\sigma +
i\gamma_5 \overrightarrow{\tau}. \overrightarrow{\pi}\big) \psi + \bar\psi
\big(i\gamma_{\mu}\partial^{\mu} - g_{\omega}\gamma_{\mu}\omega^{\mu}\big) \psi
\nonumber\\
-\frac
{1}{4}G_{\mu\nu}G^{\mu\nu}+
\frac{1}{2}m^2_{\rho}\overrightarrow{\rho_{\mu}}.\overrightarrow{\rho}^{\mu}
-\frac{1}{2}g_{\rho}\bar\psi(\overrightarrow{\rho}_{\mu}.
\overrightarrow{\tau}\gamma^{\mu}) \psi
-\frac{f}{2}\partial_\mu C \partial^\mu C,
\end{eqnarray}
\noindent  where
$F_{\mu\nu} \equiv \partial_{\mu} \omega_{\nu}
- \partial_{\nu} \omega_{\mu}$,~ $G_{\mu\nu} \equiv \partial_{\mu} \rho_{\nu}
- \partial_{\nu} \rho_{\mu}$, $\psi$ is the nucleon isospin
doublet, $\overrightarrow{\pi}$ is the pseudoscalar pion field
and $\sigma$ is the scalar field. The expectation value
$<\bar\psi\gamma_o\psi>$ is identifiable as the nucleon
number density, which we denote by $n_B$.
\par
The interactions of the scalar and
the pseudoscalar mesons with
the vector boson generates a mass for the latter spontaneously
by the Higgs mechanism. The masses for the nucleon, the scalar
meson and the vector meson are respectively given by $m =
g_{\sigma} x_o ;~~m_{\sigma} = \sqrt{2\lambda} x_o 
;~~m_{\omega} = g_{\omega} x_o,$ where $x_o$ is the vacuum
expectation value of the sigma field. C-field being
non-interacting remains massless.
\par
The equation of motion for the mean vector field specifies $\omega_o$
\begin{equation}
\omega_o = \frac{n_B}{g_{\omega}x^2}~ ,~~x =
(<\sigma^2 + \overrightarrow{\pi}^2>)^{1/2}.
\end{equation}
\noindent  The equation of motion for $\sigma$ written for
convenience in terms of $y \equiv x/x_o$ is of the form
\begin{equation}
y(1-y^2) +
\frac{c_{\sigma}c_{\omega}\gamma^2k_F^6}{18\pi^4M^2y^3}
-\frac{c_{\sigma}y\gamma}{\pi^2} \int^{k_F}_o \frac{dk
k^2}{\big(\overrightarrow k^2+M^{\star 2}\big)^{1/2}} =
0,
\end{equation}
\noindent where $m^{\star} \equiv ym$ is the effective mass of
the nucleon and $c_\sigma \equiv
g_{\sigma}^2/m_{\sigma}^2;~~~c_{\omega} \equiv
g_{\omega}^2/m_{\omega}^2$. 
\par
At high densities typical of interior of neutron stars, the
composition of matter is asymmetric nuclear matter with an
admixture of electrons. The concentrations of protons and
electrons can be determined using conditions of beta equilibrium
and electrical charge neutrality. We include the interaction due
to isospin triplet $\rho$- meson in Lagrangian for purpose of
describing neutron-rich matter. The equation of motion for
$\overrightarrow{\rho}_{\mu}$, in the mean field approximation
,where $\overrightarrow{\rho}_{\mu}$ is replaced by its uniform
value $\rho_o^3$ (here superscript 3 stands for the third
component in isospin space), gives $\rho^3_o =
\frac{g_{\rho}}{2m_\rho^2} (n_p-n_n).$
\noindent The symmetric
energy coefficient that follows from the
semi-empirical nuclear mass formula is $a_{sym} = \frac{c_{\rho}
k_F^3}{12\pi^2} + \frac{k_F^2}{6(k_F^2+m^{\star
2})^{1/2}}$, where $c_{\rho} \equiv g^2_\rho/m^2_{\rho}$ and
$k_F=(6\pi^2\rho_B/\gamma)^{1/3}$ 
($\rho_B~=~n_p+n_n$) . We take the values
of $c_{\sigma}$, $c_{\omega}$ and $c_{\rho}$ by fits of
saturation density ($0.153~fm^{-3}$), the binding energy (-16.3
MeV) and symmetric energy (32 MeV) (Moller et al., 1988) in the absence of
C-field as a first approximation. These give $ c_{\sigma} =
6.20~fm^2;~~c_{\omega} = 2.94~fm^2;~~c_{\rho} = 4.6617~fm^2.$ 
\par
The diagonal components of the conserved total stress tensor
corresponding to the Lagrangian together with the equation of
motion for the fermion field (and a mean field approximation for
the meson fields) provide the following identification for the
total energy density ($\epsilon$) and pressure ($P$) for neutron
star system : \begin{eqnarray}\varepsilon = \frac{m^2(1-y^2)^2}{8c_{\sigma}} +
\frac{\gamma^2
c_\omega ({k_p}^3+{k_n}^3)^2}{72\pi^2y^2} +\frac{\gamma^2
c_{\omega}({k_p}^3-{k_n}^3)^2}{72\pi^4y^2} \nonumber \\
+\frac{\gamma}{2\pi^2}  \sum_{n,p,e}\int_o^{k_F} dk
k^2\big(\overrightarrow{k}^2 +
m^{\star 2}\big)^{1/2} -\frac{f}{2}\dot{C}^2 \\P = -
\frac{m^2(1-y^2)^2}{8c_{\sigma}} +\frac{\gamma^2
c_\omega ({k_p}^3+{k_n}^3)^2}{72\pi^2y^2}+
\frac{\gamma^2
c_{\omega}({k_p}^3-{k_n}^3)^2}{72\pi^4y^2}\nonumber \\
+ \frac{\gamma }{6\pi^2}  \sum_{n,p,e}\int_o^{k_F} \frac{dk
k^4}{\big(\overrightarrow{k}^2 + m^{\star 2} \big)^{1/2}}
-\frac{f}{2}\dot{C}^2 .\end{eqnarray}
\noindent A specification of the coupling constants
$c_{\sigma}$, $c_{\omega}$, $c_{\rho}$ and $f$ now specifies the EOS.
\par
As of the coupling constant $f$, since it has the dimension of
$(mass)^2$, we have parameterized it in two ways like,
$f=\hat{f}m^2$ and $f=\hat{f}\frac{n_B}{m}$, where m is
typically the nucleon mass, $n_B$ is the baryon density and
$\hat{f}$ is dimensionless. The second parameterization employing a
linear proportionality between $f$ and $n_B$ is more reasonable and
physical as the coupling of C-field is likely to grow in strength with
increase in $n_B$ since creation of baryons occurs in association
with creation of C-field.
\par
The structure of a neutron star is characterized by its
gravitational mass (M) and radius (R). These gravitational mass
and radius for non-rotating neutron star are obtained by
integrating the structure equations, which describe the
hydrostatic equilibrium of degenerate stars : (Misner, Thorne \& Wheeler 
1970)
\begin{equation}\frac {dp}{dr} = - \frac{G (\rho + p/c^2) (m + 4\pi r^3 p/c^2)}
{r^2(1-2 Gm/rc^2)};~~~\frac {dm}{dr} = 4\pi r^2\rho
,\label{eq:tov2}\end{equation}
\noindent where $p$ and
$\rho(\epsilon /c^2)$ are the pressure
and total mass energy density. For a given EOS, p($\rho$), and a
given central density $\rho(r=0)=\rho_c$,  eqs.(\ref{eq:tov2}) are
integrated numerically with the boundary condition $ m(r = 0) =
0$ to give R and M.  The radius R is defined by the point 
where P $\simeq$ 0, or, equivalently, $\rho = \rho_s$, where $\rho_s$
is the density expected at the neutron star surface. The maximum total 
gravitational mass for stable configuration is then given by: 
$ M = m(R)$. The results obtained after
the inclusion of contribution of the C-field are given in table
I  and figure I. In table I, we have presented the maximum mass (M)
and the corresponding radius (R) for a more typical stable star as 
a function of central density $(\rho_c)$. It is observed from 
the figure that with increase of $\hat{f}$, the EOS becomes softer 
leading to reduction of stable neutron star gravitational mass.
\par
A comparison of the values of the radii (R) and gravitational
mass (M) for various values of dimensionless coupling parameter
$\hat{f}$ with those in the absence of C-field as obtained in an
earlier work (Sahu, Basu \& Datta 1993) (case III of Table I) reveals 
that the size and mass of stable
neutron star structures have been significantly reduced. This
can be understood on the ground that the negative energy
supplements gravitational binding and the negative pressure
reduces opposition to gravitational agregation of matter. Thus,
C-field doubly facilitates formation of compact stable neutron
star structure of smaller mass and dimension. In fact, for
$\hat{f}$ values near 0.05 and 1.5 in the two schemes of
parameterization of dimensional coupling constant $f$, the stable
neutron star mass is about 0.5$M_{\odot}$ which is well inside
the mass range of dwarf stars recently observed between $0.01$ to
1.0$M_{\odot}$ considered to be candidates for the dark
matter ( Boughn \& Uson 1995; Cottingham, Kalafatis \& Vinh Mau 1994).
Further, the value of $\hat{f}$ being of O(1)
for the mass of compact object to lie in the right range implies that
the interaction involved is strong in character as is to be expected
between the nucleon and scalar field.
\par
Apart from the above agreement, we believe that the study of the
effect and role of the C-field in various astrophysical problems
needs to be taken up in its own merit.
This provides scope to investigate the existence of explanations
alternative to those found within the ambit of the generally
accepted big bang model of the Universe.
The present work completes a small programme in that direction.
\par
It is a pleasure for us to thank Professor J. V. Narlikar for 
encouragement and suggestions. We would like to thank Professor
A. R. Prasanna for critical reading and suggetions.
One of us (PKS) would like to thank Institute of Physics, 
Bhubaneswar for facilities, where the preliminary work was started there.
\vfil
\eject
\newpage
\begin{center}
{\bf REFRENCES}
\end{center}

\noindent
Ainsworth, T. L., Baron, E., Brown, G. E., Cooperstein, J. and 
Prakash, M. 1987, {\it Nucl. Phys.}, {\bf A 464}, 740.

\noindent
Alcock, C., et al. 1993, {\it Nature}, {\bf 365}, 621.

\noindent
Arp, H. C., et al. 1990, {\it Nature}, {\bf 346}, 807.

\noindent
Aubourg, E., et al. 1993, {\it Nature}, {\bf 365}, 623.

\noindent
Boughn, S. P., and Uson, J. M. 1995, {\it Phys. Rev. Lett.}, {\bf 74}, 216.

\noindent
Cottingham, W. N., Kalafatis, D. and Vinh Mau, R. 1994, {\it Phys. Rev. Lett.}
, {\bf 73}, 1328.

\noindent
Hoyle, F., and Narlikar, J. V. 1962, {\it Proc. Roy. Soc.}, {\bf
A270}, 334.

\noindent
Jackson, A. D., Rho, M. and Krotscheck, E. 1985, {\it Nucl. Phys.},  
{\bf A407}, 495.

\noindent
Misner, C. W., Thorne, K. S. and Wheeler, J. A. 1970, {\it Gravitation}
 (San Francisco. Freeman).

\noindent
Moller, P., Myers, W. D., Swiatecki, W. J. and Treiner, J. 
1988, {\it Atomic Data Nucl. Data Tables}, {\bf 39}, 225.

\noindent
Narlikar, J. V. 1973, {\it Nature} {\bf 242}, 135.

\noindent
Narlikar, J. V. 1993, {\it Introduction to
Cosmology},  Cambridge University Press, Ch. 11.

\noindent
Pryce, M. H. L.~~~  unpublished.

\noindent
Sahu, P. K., Basu, R. and Datta, B. 1993, {\it Astrophys. J.},
{\bf 416}, 267.

\noindent
Weinberg, S. 1972, {\it Gravitation and
Cosmology: principles and applications of the general theory of
relativity}, John Wiley and Sons, Ch. 16. 
\vfil
\eject
\newpage
\begin{table}
\caption {The maximum mass (M), the corresponding
radius (R) and central density ($\rho_c$) of more typical compact 
dwarf stars for various values of dimensionless coupling parameter 
$\hat{f}$ for three cases; Case I: $f=\hat{f}m^2$, Case II: 
$f=\hat{f}n_B/m$ and Case III: $f=0$. }
\hskip 0.5 in
\begin{tabular}{ccccccc}
\hline
\multicolumn{1}{c}{$\rho_c$} &\multicolumn{1}{c}{$R$}&
\multicolumn{1}{c}{$M/M_{\odot}$}&
\multicolumn{1}{c}{$\hat{f}$}&
\multicolumn{1}{c}{$Cases$}\\
\multicolumn{1}{c}{($g~cm^{-3}$)} &\multicolumn{1}{c}{($km$)}
&\multicolumn{1}{c}{} &
\multicolumn{1}{c}{} &
\multicolumn{1}{c}{}\\
\hline
2.0$\times 10^{15}$&10.96&2.16&0.001&\\
4.0$\times 10^{15}$&7.61&1.51&0.005&I\\
7.0$\times 10^{15}$&5.94&1.19&0.01&\\
2.5$\times 10^{16}$&3.12&0.64&0.05&\\
\hline
3.0$\times 10^{15}$&9.10&1.678&0.5&\\
7.5$\times 10^{15}$&5.66&1.01&1.0&II\\
2.0$\times 10^{16}$&3.61&0.67&1.5&\\
\hline
1.5$\times 10^{15}$&13.65&2.59&0.00&III\\
\hline
\end{tabular}
\end{table}
\vfill
\eject
\newpage
\begin{figure}
\caption { Pressure and energy density curves for three cases; Case
I: $f=\hat{f}m^2$ with $\hat{f}=0.001$ (curve b) and $\hat{f}=0.05$
(curve c); Case II: $f=\hat{f}n_B/m$ with $\hat{f}=0.5$ (curve d) and
$\hat{f}=1.5$ (curve e); Case III: $f=0$ (curve a).}
\end{figure}
\vfil
\eject
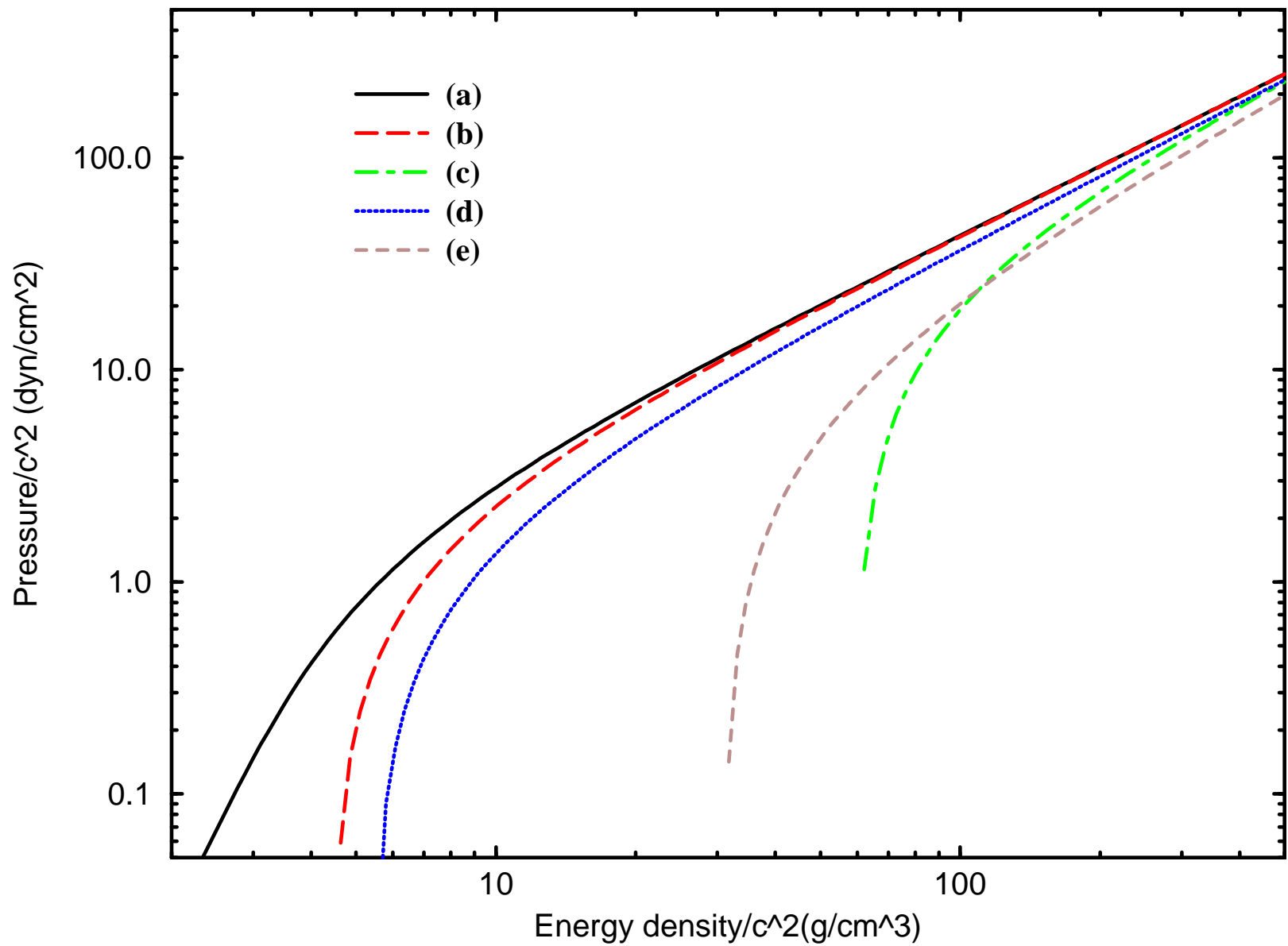
\end{document}